\documentclass[sigconf]{acmart}

\acmConference[ESEC/FSE 2019]{The 27th ACM Joint European Software Engineering Conference and Symposium on the Foundations of Software Engineering}{26--30 August, 2019}{Tallinn, Estonia}

\usepackage{amsthm}
\usepackage{algorithmicx}
\usepackage{algpseudocode}
\usepackage{algorithm}
\usepackage{xspace}
\usepackage{enumitem}
\usepackage{balance}

\newcommand{\tool}{{\sc PolyDroid}\xspace}

\newtheorem{theorem}{Theorem}[section]
\newtheorem{definition}[theorem]{Definition}
\newtheorem{remark}[theorem]{Remark}

\author{Brian Heath}
 \affiliation{University of Pennsylvania}
 \email{bheath@seas.upenn.edu}
\author{Neelay Velingker}
 \affiliation{North Penn High School}
 \email{nvelingker@gmail.com}
\author{Osbert Bastani}
 \affiliation{University of Pennsylvania}
 \email{obastani@seas.upenn.edu}
\author{Mayur Naik}
 \affiliation{University of Pennsylvania}
 \email{mhnaik@seas.upenn.edu}

\begin{document}

\title{\!\!\tool: Learning-Driven Specialization of Mobile Applications\!\!}

\begin{abstract}
The increasing prevalence of mobile apps has led to a proliferation of resource usage scenarios in which they are deployed.
This motivates the need to specialize mobile apps based on diverse and varying preferences of users.
We propose a system, called \tool, for automatically specializing mobile apps based on user preferences.
The app developer provides a number of candidate configurations, called reductions, that limit the resource usage of the original app.
The key challenge underlying \tool concerns learning the quality of user experience under different reductions.
We propose an active learning technique that requires few user experiments to determine the optimal reduction for
a given resource usage specification.
On a benchmark suite comprising 20 diverse, open-source Android apps, we demonstrate that on average, \tool obtains more than 85\% of the optimal performance using just two user experiments.
\end{abstract}

\maketitle

\begin{figure*}
\begin{tabular}{ccccc}
\includegraphics[width=0.17\textwidth]{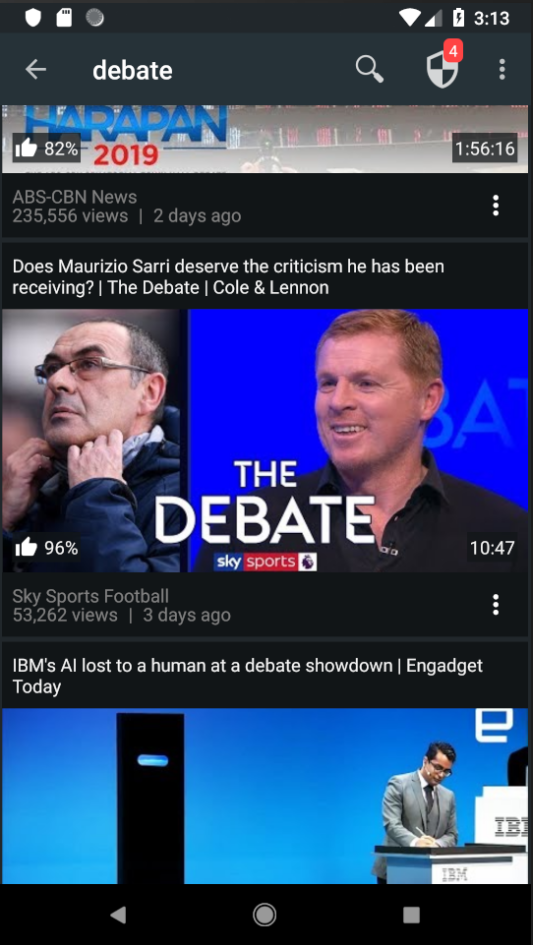} ~&~
\includegraphics[width=0.17\textwidth]{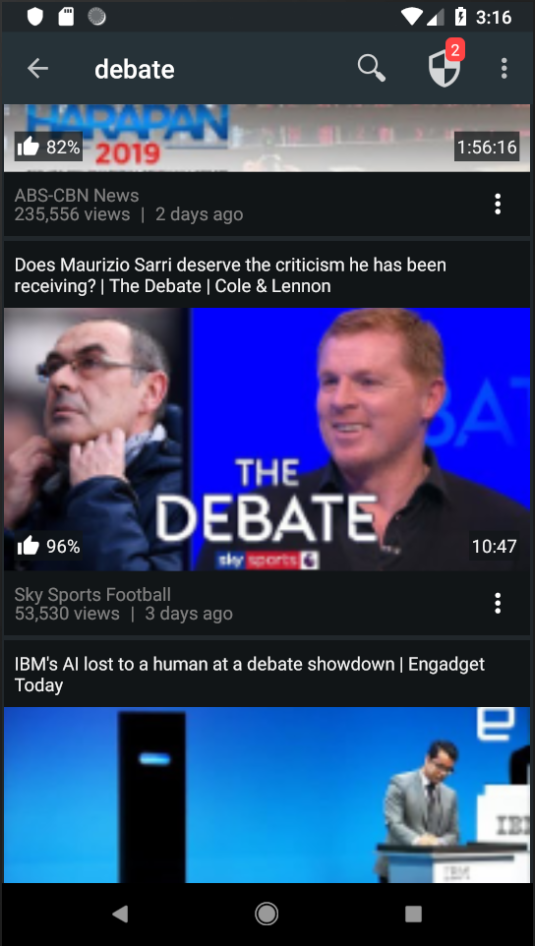} ~&~
\includegraphics[width=0.17\textwidth]{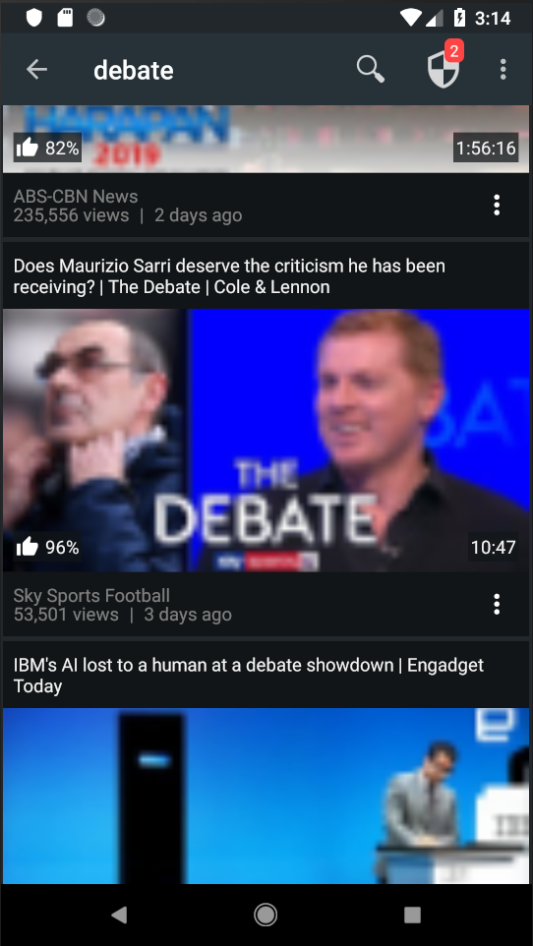} ~&~
\includegraphics[width=0.17\textwidth]{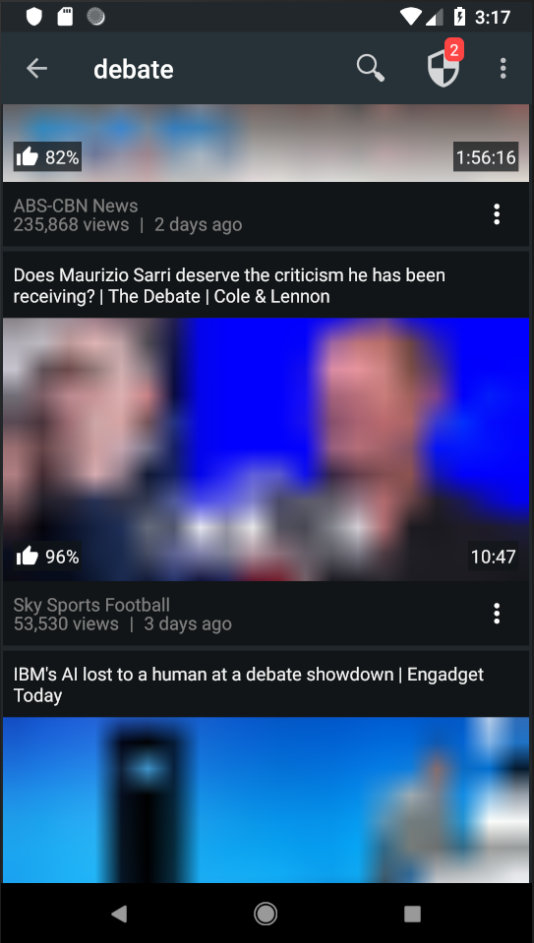} ~&~
\includegraphics[width=0.17\textwidth]{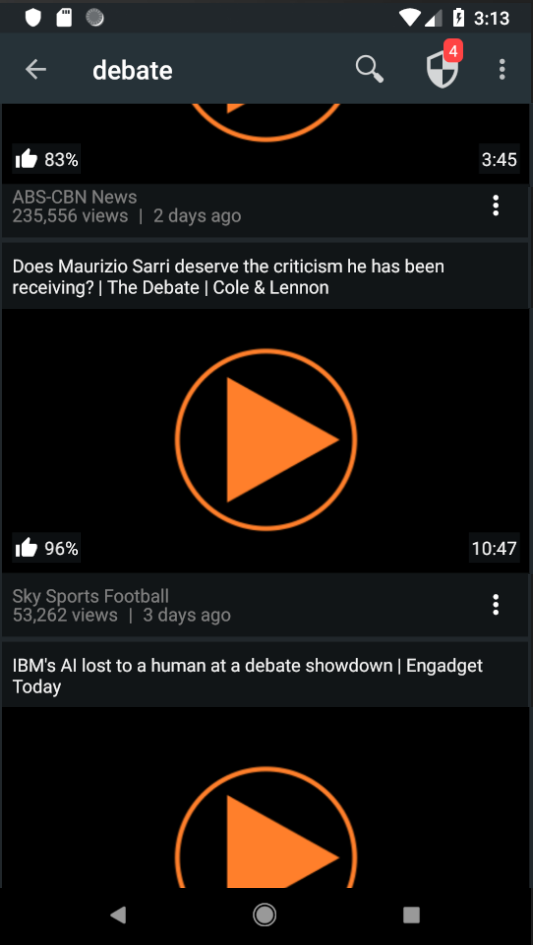} \\
original & high quality & medium quality & low quality & image removal
\end{tabular}
\vspace{-0.05in}
\caption{A screenshot from the Android app ``SkyTube'' (left-most), together with screenshots from four candidate reductions.}
\label{fig:example}
\end{figure*}

\begin{table*}
{\small
\begin{tabular}{ccccc}
\toprule
\multicolumn{1}{c}{{\bf Reduction}} & \multicolumn{1}{c}{{\bf User Experience Score (1 - 9)}} & \multicolumn{1}{c}{{\bf \% CPU Savings}} & \multicolumn{1}{c}{{\bf \% Memory Savings}} & \multicolumn{1}{c}{{\bf \% Network Data Savings}} \\
\midrule
original & 9.0 & 0.0\% & 0.0\% & 0.0\% \\
high quality & 7.4 & 0.0\% & 8.3\% & 72.0\% \\
medium quality & 4.8 & 0.0\% & 17.0\% & 88.2\% \\
low quality & 1.8 & 0.0\% & 22.4\% & 93.5\% \\
image removal & 3.1 & 0.0\% & 33.2\% & 93.7\% \\
\bottomrule
\end{tabular}}
\caption{User experience score and resource usage metrics for the original app and each candidate reduction shown in Figure~\ref{fig:example}. The user experience scores are averaged over 10 users.}
\label{tab:examplestats}
\vspace{-0.2in}
\end{table*}

\section{Introduction}
\label{sec:intro}

Mobile devices have given billions of users across the world access to web services, including many regions where access was previously unavailable. A consequence of this success is that mobile app developers must cater to a huge range of devices, from expensive, high-end devices that are nearly as powerful as more traditional computers, to inexpensive, low-end devices that have limited computing resources. Even a single user typically has widely varying preferences---e.g., depending on whether they are at home or work, traveling, or have limited access to the cellular network.

Thus, there is a pressing need for tools to help developers specialize their app based on resource usage preferences provided by the user. Existing tooling provides limited support for such specialization---for example, Android devices offer ``battery saver'' and
``data saver'' modes that reduce battery usage and network data usage, respectively, to allow users to dynamically configure apps based on their current preferences regarding resource usage. However, beyond general rules enforced by the operating system (e.g., turning off background GPS usage),
it is up to the developer to determine how to specialize the app. That is, the developer must manually specify how to modify the behavior of their app when the user changes their resource usage preferences.

A key challenge in automatically specializing mobile apps is that different configuration options often affect user experience. For example, one way to reduce network data usage is to reduce the resolution of the images downloaded by an app. However, how the reduction in quality affects user experience depends heavily on the context in which the image appears---e.g., reducing the quality of images of restaurant dishes in a food review app may render the app useless, whereas reducing the quality of background images in a weather app may have little or no impact on usability. Similarly, an effective way to reduce CPU usage is to eliminate animations when transitioning between activities, but doing so may substantially reduce user experience.

We propose a system, called \tool, for automatically specializing mobile apps based on user preferences. To use \tool, the developer simply provides a number of candidate configurations, called \emph{reductions}, along with the original app, as well as a test script for each reduction that exercises the behaviors in that reduction that differ from the original app. Reductions and test scripts can also be automatically generated using program analysis. Then, whenever a user desires to limit the resource usage of the app, \tool selects the reduction that optimizes a combination of (i)~the resource usage of the reduction (depending on the user's preferences), and (ii) the user's experience of the reduction.
In particular, when using \tool, the developer does not have to specify either the resource usage of different reductions, or their impact on user experience (which is best dictated by the user). Thus, \tool effectively separates the concern of optimizing this tradeoff from the
implementation of the mobile app.

To determine the resource usage of different reductions, \tool runs each reduction in emulation and records its resource usage.\footnote{This strategy is reasonable since emulators for Android apps have very sophisticated tools for measuring resource usage; if desired, resource usage could be measured on real devices as well.}
The more challenging problem is measuring the quality of the user experience for different candidate reductions. In particular, doing so requires running user experiments, which can be expensive at large scale. The key contribution of \tool is to leverage \emph{active learning} to substantially
reduce the number of user experiments. \tool uses an active learning strategy based on the Thompson sampling algorithm~\cite{chapelle2011empirical,lattimore2018bandit}. In addition, \tool uses historical data to estimate a Bayesian prior to guide sampling. Using this strategy, \tool can produce sensible results (e.g., about 68\% of optimal) even with \emph{zero} experiments on the current app (i.e., based on data from other apps alone). Furthermore, running just two of the total possible user experiments (about 11 on average) is sufficient to get more than 85\% of optimal performance.

We have implemented \tool for the Android platform and evaluate it using a benchmark suite comprising 20 diverse, open-source apps. We show that by leveraging active learning, \tool can compute good reductions based on very few user experiments. Furthermore, we show how \tool can be used to \emph{personalize}
the choice of reduction by directly querying the end user (as opposed to, for instance, an online survey of hired users).

In summary, our work makes the following contributions:
\begin{itemize}[leftmargin=*,topsep=4pt]
\item We formulate the problem of determining the optimal reduction for a user-provided resource usage specification (Section~\ref{sec:problem}).
\item We provide an algorithm for computing the optimal reduction, as well as a variant that uses active learning to minimize the number of user experiments needed (Section~\ref{sec:algo}).
\item We have implemented our approach in a tool called \tool (Section~\ref{sec:impl}) and show that it can compute good reductions using just a small number of user experiments (Section~\ref{sec:eval}).
\end{itemize}

\begin{figure*}
\vspace{-0.05in}
\includegraphics[width=0.8\textwidth]{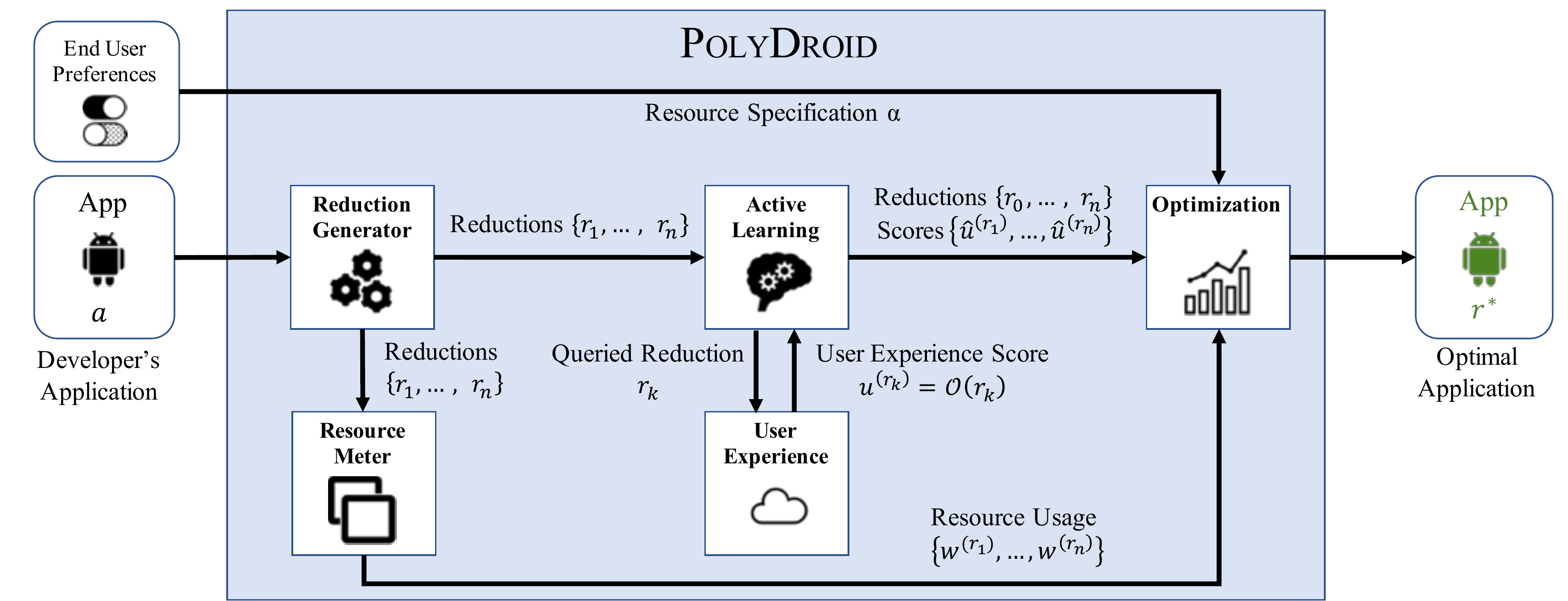}
\caption{An overview of our proposed tool \tool. Given a mobile app and resource usage preferences provided by the end user, \tool computes the optimal reduction of the app to give to the end user.}
\label{fig:system}
\end{figure*}

\vspace{-0.1in}
\section{Motivating Example}
\label{sec:example}

Suppose that Alice develops a new Android app for watching and sharing videos, and wants it to be accessible to a wide variety of users. Using the current tools available, Alice must manually specify how to modify her app to suit every different resource usage specification that an end user provides. For example, suppose that Bob downloads Alice's app. When Bob is traveling, he may want to conserve network data usage, but he wants to continue to use Alice's app to keep track of the news. While the Android OS allows Bob to specify this preference, Alice would have to implement program logic that encodes how to reduce network data usage. For example, Alice can reduce the quality of downloaded images.

A key challenge is that Alice has to reason about how different modifications to her app affect the user experience of her app. For example, suppose that Alice has implemented the candidate modifications to her app shown in Figure~\ref{fig:example}. We refer to each modified version of the app as a \emph{reduction}. These reductions modify the original app by reducing the quality of images in the app by different degrees, or even removing the images altogether. For each of these candidate reductions, Alice has to understand whether the user would be satisfied with the app, either through user experiments or based on her intuition. Then, Alice must combine her estimate of the user experience with every possible resource usage specification that Bob may provide to decide which reduction to use in each instance. As a consequence, to ensure that Alice's workload is reasonable, the Android OS currently only provides support for a very limited number of resource usage specifications.

In our approach, Alice only needs to provide our tool \tool with a set of candidate reductions to be considered, together with a test script exercising the modified behaviors in each reduction compared to the original app. In particular, she does not need to specify anything about the tradeoff between resource consumption and user experience of each reduction. Instead, \tool automatically measures these quantities for each reduction. Then, whenever Bob provides a new resource usage specification, \tool uses this information to choose the reduction that best suits Bob's needs. We describe how \tool does so in more detail using a series of scenarios. Also, we give an overview of the architecture of \tool in Figure~\ref{fig:system}.

\vspace{-0.05in}
\paragraph{{\bf Learning from user experiments.}}

When Bob first installs Alice's app, he wants to customize it to run smoothly on his low-end device. In particular, his device has limited memory, and his phone plan has limits on his network data usage. Thus, Bob wants reduce the memory usage and network data usage of Alice's app if possible,
though not at the cost of a significant reduction in user experience. \linebreak To do so, Bob provides a specification
\begin{align*}
\lambda_{\text{low-end}}&=0.5 \\
\alpha_{\text{low-end}}&=\{\text{CPU}:~0.0,~\text{memory}:~0.5,~\text{network}:~0.5\}.
\end{align*}
This specification says that memory usage is half as important as user experience---e.g., decreasing both memory usage by 10\% and network data usage by 10\% is worth reducing user experience by at most 20\%. In particular, the portion $\alpha_{\text{low-end}}$ specifies the relative importance of different resources---e.g., CPU, memory, and network data (so this specification says that Bob cares equally about memory usage and network data usage, but not at all CPU usage), and the portion $\lambda_{\text{low-end}}$ specifies that overall tradeoff between resource savings and user experience (so this specification says that saving resources is $0.5\times$ as important to Bob as user experience).

Since Alice's app has been available for a while, \tool has already run user experiments to determine the user experience scores of each of the available reductions (an integer between 1 and 9). \tool has also run each of the reductions in emulation to determine their peak CPU usage, memory usage, and network data usage. All of these data are shown in Table~\ref{tab:examplestats}.\footnote{Note that CPU usage is not reduced in any of these reductions. Other modifications can be used to reduce CPU usage; see Section~\ref{sec:eval} for details.}

As a consequence, \tool can immediately determine which reduction optimizes the tradeoff between resource consumption and user experience, based on the specification that Bob provided. In particular, \tool selects the high quality reduction in Figure~\ref{fig:example}; this reduction trades a small decrease in user experience for a large decrease in network data usage and memory usage.

\begin{remark}
\rm
In this paper, we focus on the usage of CPU, memory, and network data, since the consumption of these resources can effectively be estimated in an emulator. \tool can easily be extended to handle battery usage by measuring the battery consumption of different reductions of an app on a real device.
\end{remark}

\begin{remark}
\rm
In practice, we envision that Bob would provide the specification $\alpha_{\text{low-end}}$ similarly to how Android lets users activate a ``battery saver'' or ``data saver'' mode. For example, the interface may allow Bob to drag different sliders letting him specify the degree to which he cares about CPU usage, memory usage, and network data usage. Then, the Android OS would automatically customize the app to work with the given resource consumption levels. If desired, Bob would also be able to tune these values for specific apps.
\end{remark}

\paragraph{{\bf Learning from a few user experiments.}}

A few months later, Alice releases an updated version of her app. Bob is one of the first users to install the updated app. Bob provides the same resource usage specification $\alpha_{\text{low-end}}$ as before. Since the app has not changed very much, re-running the user experiments for all the reductions is unnecessary and a waste of resources. Thus, \tool uses active learning to select a small number of promising reductions for which to run user experiments. These choices are guided by historical data on other apps, including the previous versions of Alice's app. As the app has not changed substantially, \tool selects the same reduction as before.

\paragraph{{\bf Learning from zero user experiments.}}

While traveling, Bob wants to limit his usage of network data to avoid incurring high roaming charges. Thus, Bob provides a new specification
\begin{align*}
\lambda_{\text{travel}}&=1.0 \\
\alpha_{\text{travel}}&=\{\text{CPU}:~0.0,~\text{memory}:~0.25,~\text{network}:~0.75\},
\end{align*}
which indicates that Bob is willing to substantially reduce user experience in exchange for reduced network data usage. Since Bob needs \tool to choose a reduction quickly, there may be no time to run additional user experiments. Thus, \tool relies on historical data (along with the few user experiments it has already run for reductions of Alice's apps). More precisely, \tool uses features of historical reductions and the outcomes of user experiments for these reductions in conjunction with machine learning to predict user experience measures for the reductions of Alice's app. In this case, \tool selects the image removal reduction in Figure~\ref{fig:example}, which substantially reduces network data usage but also substantially diminishes user experience.

\paragraph{{\bf Personalized reductions.}}

Bob does not like the reduction---unlike the average user, who prefers to see no images at all than to seeing low quality images, Bob prefers to see images even if they are low quality. In this case, Bob can ask \tool to \emph{personalize} the chosen reduction to his needs. To do so, \tool directly asks Bob to respond to user queries to elicit his personal preferences. \tool continues to query Bob until it chooses a reduction that Bob finds satisfactory. For example, \tool may ask Bob to score the image removal reduction in Figure~\ref{fig:example}, since it thinks this reduction is a promising choice. However, Bob gives this reduction a user experience score of 1 (the lowest possible), so \tool learns that it is not a good choice. With this new information, \tool asks Bob to score the low quality reduction, since it is now very uncertain about the score for this reduction. Bob assigns this reduction a user experience score of 4. Now that \tool has exhausted its query budget, it uses the information Bob has provided so far to choose a reduction. In particular, \tool selects the low quality reduction. In this setting, it is particularly important to use active learning to minimize the number of queries needed---otherwise, Bob may have to spend a large amount of time responding to queries before \tool returns a satisfactory reduction.

\section{Problem Formulation}
\label{sec:problem}

In this section, we begin by introducing the problem of computing optimal reductions for a user-provided specification indicating the desired usage level for each resource. Then, we describe the problem of reducing the number of queries made to evaluate user experience when computing the optimal reduction.

\paragraph{{\bf Optimal reduction.}}

Consider a mobile app $a$, together with a set of \emph{reductions} $\mathcal{R}(a)$, where each reduction $r\in\mathcal{R}(a)$ is a modified version of $a$ designed to consume fewer resources than $a$ (e.g., in terms of CPU usage, memory usage, or network data usage), possibly at the expense of degraded user experience. Reductions may be provided by the developer, or may be constructed by automatically modifying the given app. 

We assume that each reduction $r$ is associated with a user experience score $u^{(r)}\in\mathbb{R}$ and a vector $w^{(r)}\in\mathbb{R}^d$, where $w^{(r)}_i$ measures the level of usage of resource $i$. In particular, $w^{(r)}_i\in[0,1]$ is the reduction in usage of resource $i$ for $r$ compared to the original app $a$. Thus, $w^{(r)}_i=0$ means that $r$ does not provide any resource savings compared to $a$, whereas $w^{(r)}_i=1$ means that $r$ has zero usage of resource $i$. Note that larger $w^{(r)}_i$ is preferable. Similarly, $u^{(r)}$ is normalized to $[0,1]$. In particular, $u^{(r)}$ measures the fraction of user experience retained in $r$ compared to the original app $a$. Thus, $u^{(r)}=1$ means that the user experience of $r$ is as good as that of $a$, whereas $u^{(r)}=0$ means $r$ is unusable. As before, larger $u^{(r)}$ is preferable.

Next, we assume that the user specifies preferences as follows.
\begin{definition}
\rm
A \emph{resource usage specification} is a pair $(\lambda,\alpha)\in\mathbb{R}\times\mathbb{R}^d$, such that $\alpha_i\in[0,1]$ for each $i\in\{1,...,d\}$, and $\sum_{i=1}^d\alpha_i=1$. In particular, $\lambda$ specifies the overall importance of resource usage relative to user experience, and $\alpha_i$ specifies the relative importance of resource $i$ for each $i\in\{1,...,d\}$.
\end{definition}
Then, given an app $a$ and a (user-provided) resource specification $(\lambda,\alpha)$, our goal is to find the app that optimizes the tradeoff between resource consumption and user experience.
\begin{definition}
\rm
The \emph{resource-aware user experience} $J(r;\lambda,\alpha)$ is
\begin{align}
\label{eqn:optr}
r^*&=\operatorname*{\arg\max}_{r\in\mathcal{R}(a)}J(r;\lambda,\alpha) \\
J(r;\lambda,\alpha)&=u^{(r)}+\lambda\langle\alpha,w^{(r)}\rangle, \nonumber
\end{align}
where $\langle\cdot,\cdot\rangle$ is the inner product on $\mathbb{R}^d$. We refer to $r^*$ as the \emph{optimal reduction}, and $J(r^*;\lambda,\alpha)$ as the \emph{optimal objective value}, for app $a$ and specification $(\lambda,\alpha)$.
\end{definition}

\paragraph{{\bf Querying user experience.}}

When we know the user experiences score $u^{(r)}$ for each reduction $r\in\mathcal{R}_a$, then solving (\ref{eqn:optr}) is straightforward---given $(\lambda,\alpha)$, we compute $J(r;\lambda,\alpha)$ for each $r\in\mathcal{R}(a)$, and choose $r$ that maximizes $J(r;\lambda,\alpha)$. However, a key challenge is that the scores $u^{(r)}$ are initially unknown.

To determine $u^{(r)}$ for a given reduction $r$, we can query users to assess the usability of $r$ compared to the original app $a$. We assume that the app developer provides test scripts $t^{(r)}$ for each $r\in\mathcal{R}(a)$. Then, we run the test script on each reduction in an emulator while recording the emulator screen. Given the recordings we evaluate $u^{(r)}$ as follows: we show the user both the recording for the original app $a$ and the recording for the reduction $r$ side by side, and ask them to rate the quality of the reduction according to a scale from 1 to 9, where 1 means that $r$ is completely unusable and 9 means that $r$ is indistinguishable from $a$. We describe our implementation of this user experiment in more detail in Section~\ref{sec:userqueries}. Finally, we use $u^{(r)}=\mathcal{O}(r)$ to denote the response to a query on reduction $r$.

\begin{remark}
\rm
When querying the user, we are asking them to evaluate only the user experience of the reduction, ignoring resource consumption issues; issues such as latency should be captured by the resource usage specification $(\lambda,\alpha)$. In this paper, we assume that $(\lambda,\alpha)$ is known. It may also be possible (and useful) to infer $(\lambda,\alpha)$ from repeated interactions with the user; we leave this possibility to future work.
\end{remark}

\paragraph{{\bf Measuring resource usage.}}

We determine resource usage $w^{(r)}$ by running the app (e.g., in emulation) and monitoring the resource consumption. We discuss implementation details in Section~\ref{sec:resourceconsumption}.

\paragraph{{\bf Active learning.}}

Querying each reduction of each app can be very costly---there are millions of apps on the app store, each of which may come with dozens of candidate reductions, and a query for each reduction can involve asking multiple questions to multiple users. To alleviate this problem, we propose an approach where we actively query users for their preference $u^{(r)}$ of certain reductions.

In particular, upon providing a resource usage specification, our algorithm issues queries on a small number $B\in\mathbb{N}$ (e.g., $B=1$ or $B=2$) of actively chosen reductions $r\in\mathcal{R}_a$; we call $B$ the \emph{query budget}. Our goal is to use these samples to compute an estimate $\hat{r}$
that is close in quality to the optimal reduction $r^*$ in equation (\ref{eqn:optr}). More precisely, the goal of our algorithm is to return a reduction $\hat{r}$ such that $J(\hat{r};\lambda,\alpha)$ is close to the optimal objective value $J(r^*;\lambda,\alpha)$.

\section{Algorithm}
\label{sec:algo}

We begin by describing how \tool optimizes the reduction $r\in\mathcal{R}(a)$ of an app $a$ given a user-provided specification $(\lambda,\alpha)$. In particular, we describe how \tool performs user experiments to determine the user experience of different reductions, leveraging active learning to minimize the number of user queries. We also describe two variants of \tool---one which queries the user on every reduction, and one which queries the user on zero reductions. These variants represent extremal use cases of \tool.

\subsection{Reduction Optimization Algorithm}

\begin{algorithm}[t]
\begin{algorithmic}
\Procedure{Polydroid}{App $a$, Specification $(\lambda,\alpha)$, Query budget $B\in\mathbb{N}$, Historical data $\mathcal{H}$}
\State $\mathcal{R}\gets\mathcal{R}(a)$
\State $\mathcal{D}\gets\varnothing$
\State $\theta_0\gets$~\Call{LearnPrior}{$\mathcal{H}$}
\ForAll{$i\in\{1,...,B\}$}
\State $r\gets$~\Call{ThompsonSample}{$(\lambda,\alpha)$, $\mathcal{R}$, $\mathcal{D}$, $\theta_0$}
\State $u^{(r)}\gets\mathcal{O}(r)$
\State $\mathcal{R}\gets\mathcal{R}\setminus\{r\}$
\State $\mathcal{D}\gets\mathcal{D}\cup\{(r,u^{(r)})\}$
\EndFor
\State $\hat{r}\gets$~\Call{OptimizeReduction}{$(\lambda,\alpha)$, $\mathcal{R}$, $\mathcal{D}$, $\theta_0$}
\State \Return $\hat{r}$
\EndProcedure
\Procedure{LearnPrior}{Historical data $\mathcal{H}$}
\State \Return $\operatorname*{\arg\max}_{\theta\in\Theta_0}\ell(\theta;\mathcal{H})$
\EndProcedure
\Procedure{ThompsonSample}{Specification $(\lambda,\alpha)$, Un-queried reductions $\mathcal{R}$, Current data $\mathcal{D}$, Prior parameters $\theta_0$}
\State $\tilde{\theta}_a\sim p(\cdot;\theta_0,\mathcal{D})$
\ForAll{$r\in\mathcal{R}$}
\State $\hat{u}^{(r)}\gets f(r;\tilde{\theta}_a)$
\EndFor
\State \Return $\operatorname*{\arg\max}_{r\in\mathcal{R}}\{\hat{u}^{(r)}+\lambda\langle\alpha,w^{(r)}\rangle\}$
\EndProcedure
\Procedure{OptimizeReduction}{Specification $(\lambda,\alpha)$, \!Un-queried reductions $\mathcal{R}$, Current data $\mathcal{D}$, Prior parameters $\theta_0$}
\State $\hat{\theta}_a\gets\operatorname*{\arg\max}_{\theta\in\Theta}p(\theta;\theta_0,\mathcal{D})$
\ForAll{$r\in\mathcal{R}$}
\State $\hat{u}^{(r)}\gets f(r;\hat{\theta}_a)$
\EndFor
\ForAll{$(r,u^{(r)})\in\mathcal{D}$}
\State $\hat{u}^{(r)}\gets u^{(r)}$
\EndFor
\State \Return $\operatorname*{\arg\max}_{r\in\mathcal{R}(a)}\{\hat{u}^{(r)}+\lambda\langle\alpha,w^{(r)}\rangle\}$
\EndProcedure
\end{algorithmic}
\caption{The algorithm used to compute good reductions.}
\label{alg:online}
\end{algorithm}

For many apps $a$---e.g., the long tail of unpopular apps with many configurations, or apps that are frequently updated---it may be prohibitively expensive to query every reduction $r\in\mathcal{R}(a)$. Similarly, a user may want to provide their own user experience scores to personalize the optimization problem, but may be unwilling to respond to dozens or even hundreds of queries.

To handle these settings, \tool leverages active learning to minimize the number of queries made to the user. At a high level, the active learning algorithm consists of two components: (i) a Thompson sampling strategy to actively select which reduction to query next, and (ii) a Bayesian prior over the user experience scores of different reductions (learned from historical data) to guide Thompson sampling. Finally, \tool uses the results from the active learning algorithm to optimize the reduction of the app $a$. We discuss each of these components in detail below.

\paragraph{{\bf Thompson sampling for active learning.}}

First, given an app $a$, our algorithm assumes that the user experience score $u^{(r)}$ of a reduction $r\in\mathcal{R}(a)$ is modeled as a parametric function
\begin{align*}
u^{(r)}&=f(r;\theta_a)
\end{align*}
where $\theta_a\in\Theta$ are unknown parameters specific to $a$. We describe the model $f(r;\theta_a)$ we use in our implementation of \tool in Section~\ref{sec:impl}.

\tool uses active learning to estimate $\theta_a$, and then chooses the best app according to the machine learning model $f(r;\theta_a)$. In particular, an active learning algorithm in our setting should iteratively select reductions $r\in\mathcal{R}(a)$, query $u^{(r)}=\mathcal{O}(r)$, and then incorporate this information to help select the next reduction to query. Recall from Section~\ref{sec:problem} that the goal is to select an app with near-optimal objective $J(r^*;\lambda,\alpha)$ for a given query budget $B\in\mathbb{N}$.

The key challenge in our setting is that we need to balance the so-called exploration-exploitation tradeoff~\cite{lattimore2018bandit}. In particular, we need to focus on reductions $r$ that are most likely to achieve high objective value
\begin{align*}
J(r;\lambda,\alpha)&=f(r;\theta_a)+\lambda\langle\alpha,w^{(r)}\rangle
\end{align*}
according to our model (exploitation), while simultaneously trying reductions with lower objective value but high variance to avoid the possiblity of missing out on a good reduction (exploration). This problem is a special case of active learning known as bandit learning~\cite{lattimore2018bandit}.

The Thompson sampling algorithm is an effective approach for solving bandit learning problems. To apply Thompson sampling to our setting, we first assume that $\theta_a$ has prior distribution
\begin{align*}
\theta_a\sim p(\cdot;\theta_0),
\end{align*}
where $\theta_0\in\Theta_0$ are parameters of the prior. As we discuss below, an effective way to choose $\theta_0$ is to estimate it based on historical data. We describe the model $p(\cdot;\theta_0)$ we use in our implementation of \tool in Section~\ref{sec:impl}.

Then, Thompson sampling iteratively performs the following steps: (i) query a point $\tilde{\theta}_a\sim p(\cdot;\theta_0,\mathcal{D})$, where $p(\cdot;\theta_0,\mathcal{D})$ is the posterior probability distribution over $\theta_a$ given the data $\mathcal{D}=\{(r,u^{(r)})\}$ observed so far, (ii) pretend like our sample $\tilde{\theta}_a$ are the true parameters, select the optimal reduction $J(r;\lambda,\alpha)$ according to $\tilde{\theta}_a$, and (iii) query $u^{(r)}=\mathcal{O}(r)$. In Algorithm~\ref{alg:online}, the Thompson sampling subroutine for choosing which reduction to query is the subroutine \textsc{ThompsonSample}.

\paragraph{{\bf Learning a prior.}}

Even using Thompson sampling, our algorithm may require many samples to learn. In general, an effective way to speeding up active learning is to use a good prior distribution---i.e., to choose the parameters $\theta_0$ of the prior $p(\cdot;\theta_0)$ in a way that quickly guides the active learning algorithm towards more promising solutions. In particular, our algorithm uses historical data $\mathcal{H}=\{(a_t,r_t,u^{(r_t)})\}_{t=0}^{T-1}$ collected from past apps $a_t$ and user queries $u^{(r_t)}=\mathcal{O}(r_t)$ for $r_t\in\mathcal{R}(a_t)$ to guide the active learning.
\footnote{Note that the historical data may not be representative of the current app $a$---for example, if images are crucial to the functionality of $a$, then reductions $r\in\mathcal{R}(a)$ that degrade the quality of images in $a$ may not be suitable, even if they are suitable for most apps in $\mathcal{H}$. Thus, we can improve performance beyond the historical data by actively labeling reductions $r\in\mathcal{R}(a)$ specific to $a$.}
To do so, our algorithm uses maximum-likelihood estimation---it chooses $\theta_0$ to maximize the likelihood of observing the historical data $\mathcal{H}$ given that $\theta_0$ are the true parameters, i.e.,
\begin{align*}
\theta_0&=\operatorname*{\arg\max}_{\theta\in\Theta_0}\ell(\theta;\mathcal{H}) \\
\ell(\theta;\mathcal{H})&=p(\mathcal{H};\theta).
\end{align*}
We describe the model $p(\mathcal{H};\theta)$ we use in our implementation of \tool in Section~\ref{sec:impl}. In Algorithm~\ref{alg:online}, the \textsc{LearnPrior} subroutine learns the parameters of a prior based on historical data.

\paragraph{{\bf Optimizing the reduction.}}

Finally, \tool uses the results from the queries $u^{(r)}=\mathcal{O}(r)$ to optimize the reduction $r\in\mathcal{R}(a)$ for the given app $a$. First, it computes the maximum likelihood parameters
\begin{align*}
\hat{\theta}_a=\operatorname*{\arg\max}_{\theta\in\Theta}p(\theta;\theta_0,\mathcal{D}),
\end{align*}
according to our machine learning model of user experience. Then, it uses $\hat{\theta}_a$ to predict the user experience for reductions where the user was not queried. For apps where the user was queried, \tool simply uses the response $u^{(r)}=\mathcal{O}(r)$. Together, we have
\begin{align*}
\hat{u}^{(r)}=
\begin{cases}
f(r;\hat{\theta}_a)&\text{if}~r\in\mathcal{R} \\
u^{(r)}&\text{otherwise},
\end{cases}
\end{align*}
where $\mathcal{R}\subseteq\mathcal{R}(a)$ is the subset of reductions for which the user has not been queried. Finally, \tool returns the reduction
\begin{align*}
\hat{r}=\operatorname*{\arg\max}_{r\in\mathcal{R}(a)}\{\hat{u}^{(r)}+\lambda\langle\alpha,w^{(r)}\rangle\}
\end{align*}
that is optimal according to the estimated user experience scores $\hat{u}^{(r)}$. In Algorithm~\ref{alg:online}, \textsc{OptimizeReduction} computes $\hat{\theta}_a$, $\hat{u}^{(r)}$, and $\hat{r}$.

\subsection{Other Variants of \tool}

\paragraph{{\bf Offline Variant.}}

For some apps $a$---e.g., very popular apps or apps that have few available configurations---it makes sense to query the user experience score $u^{(r)}$ for every reduction $r\in\mathcal{R}(a)$ ahead of time. Then, given a resource specification $(\lambda,\alpha)$, \tool simply returns the reduction $r$ with the highest objective value. This variant of \tool corresponds to running Algorithm~\ref{alg:online} with a query budget of $B=|\mathcal{R}(a)|$---i.e., \tool queries $u^{(r)}=\mathcal{O}(r)$ for every reduction $r\in\mathcal{R}(a)$.

\paragraph{{\bf Zero user experiments variant.}}

In some situations---e.g., when a reduction is urgently needed, but no user experience data is available---it is desirable to obtain a reduction without user queries. In these situations, \tool can rely solely on historical data to choose the optimal reduction. This variant of \tool corresponds to running Algorithm~\ref{alg:online} with a query budget of $B=0$.

\begin{figure}
\includegraphics[width=\linewidth]{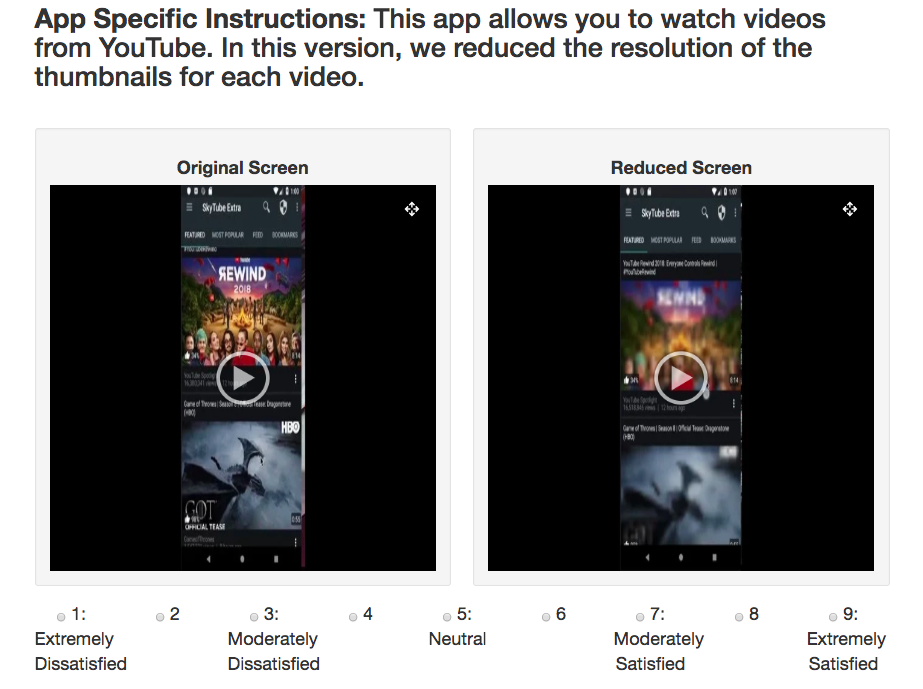}
\vspace{-0.25in}
\caption{Each version within the survey has three components. On the top, a message orients the participant to the app and the reduction. In the middle, the pair of videos has a sequence from the original version on the left, and the reduced version on the right. Lastly, there are nine buttons where the user selects their satisfaction with the reduction.}
\label{fig:survey}
\end{figure}

\vspace{-0.1in}
\section{Implementation}
\label{sec:impl}

We discuss several aspects of our implementation---in particular, how we query the user to obtain $u^{(r)}$, how we estimate the resource usage of each reduction, and some design choices in our Thompson sampling algorithm.

\subsection{User Queries}
\label{sec:userqueries}

To perform a user query $\mathcal{O}(r)$ on a reduction $r\in\mathcal{R}(a)$ for app $a$, we first run the test scripts provided by the developer on both $r$ and on $a$ and records the sequences of interactions. We show these two recordings side-by-side to the user, and ask them to watch each recording. In addition to this pair of videos, each user is shown instructions describing with the purpose of the app, the specific role of the activity within the app, and the modification that was made to the app. Then, we ask each user to score the quality of the reduction on a scale from 1 (extremely dissatisfied) to 9 (extremely satisfied), with 5 representing a neutral opinion. A single reduction within the survey can be viewed in Figure \ref{fig:survey}. The user response, normalized to $[0,1]$, is our score $u^{(r)}$. Below, we discuss a few details in our implementation.  
\paragraph{{\bf Averaging over multiple users.}}

Typically, our queries are made to paid users such as Amazon Mechanical Turk workers. In these cases, we average each query over 10 users to obtain the average user experience. Using this approach, the end user does not have to answer any queries regarding user experience. Optionally, we can directly query the end user to obtain a personalized user experience score.

\begin{table}
\caption{Summary of reductions across the 20 apps. ``\# Apps'' is the number of apps where a reduction of the given type is applicable, and ``\# Reductions'' is the total number of reductions of the given type that we constructed.}
\vspace{-0.1in}
{\small
\begin{tabular}{@{}ccc@{}}\toprule
\textbf{Reduction Type} & \textbf{\# Apps} & \textbf{\# Reductions} \\\midrule
Image Removal	&	16	&	36\\
Image Resolution $\rightarrow$ 400px $\times$ 400px	&	4	&	5\\
Image Resolution $\rightarrow$ 200px $\times$ 200px 	&	6	&	9\\
Image Resolution $\rightarrow$ 100px $\times$ 100px	&	8	&	13\\
Image Resolution $\rightarrow$\ \ 50px  $\times$ 50px\ \	&	9	&	15\\
Image Resolution $\rightarrow$\ \ 20px  $\times$ 20px\ \	&	9	&	15\\
Transition Removal	&	8	&	18\\
Image \& Transition Reduction	&	4	&	28\\\midrule
\textbf{Total}	&	20	&	139	\\\bottomrule
\end{tabular}}
\vspace{-0.05in}
\label{tbl:features}
\end{table}

\paragraph{{\bf Per-activity scores.}}

To simplify the comparison that the user needs to make, we actually query the modification to each view $v\in\mathcal{V}(r)$ in $r$ independently, where $\mathcal{V}(r)$ is the set of views in $r$. By view, we mean a single screen of the app---e.g., an activity in the case of Android apps.
\footnote{We assume that the reduction $r$ has the same activities as the original app $a$.}
More precisely, we assume that the user experience can be decomposed as
\begin{align*}
u^{(r)}=\sum_{v\in\mathcal{V}(r)}u^{(r,v)},
\end{align*}
where $u^{(r,v)}$ is the user experience for view $v$ in reduction $r$. Then, we estimate the $u^{(r,v)}$ independently, and sum them together to compute $u^{(r)}$. By doing so, we can reduce the amount of information we need to show the user at each question, since $u^{(r,v)}$ refers to a much smaller portion of the app than $u^{(r)}$. For the kinds of reductions we consider, this decomposition works well, but in general, different granularities of decompositions may be needed for different kinds of reductions.

\begin{figure}
\includegraphics[width=\linewidth]{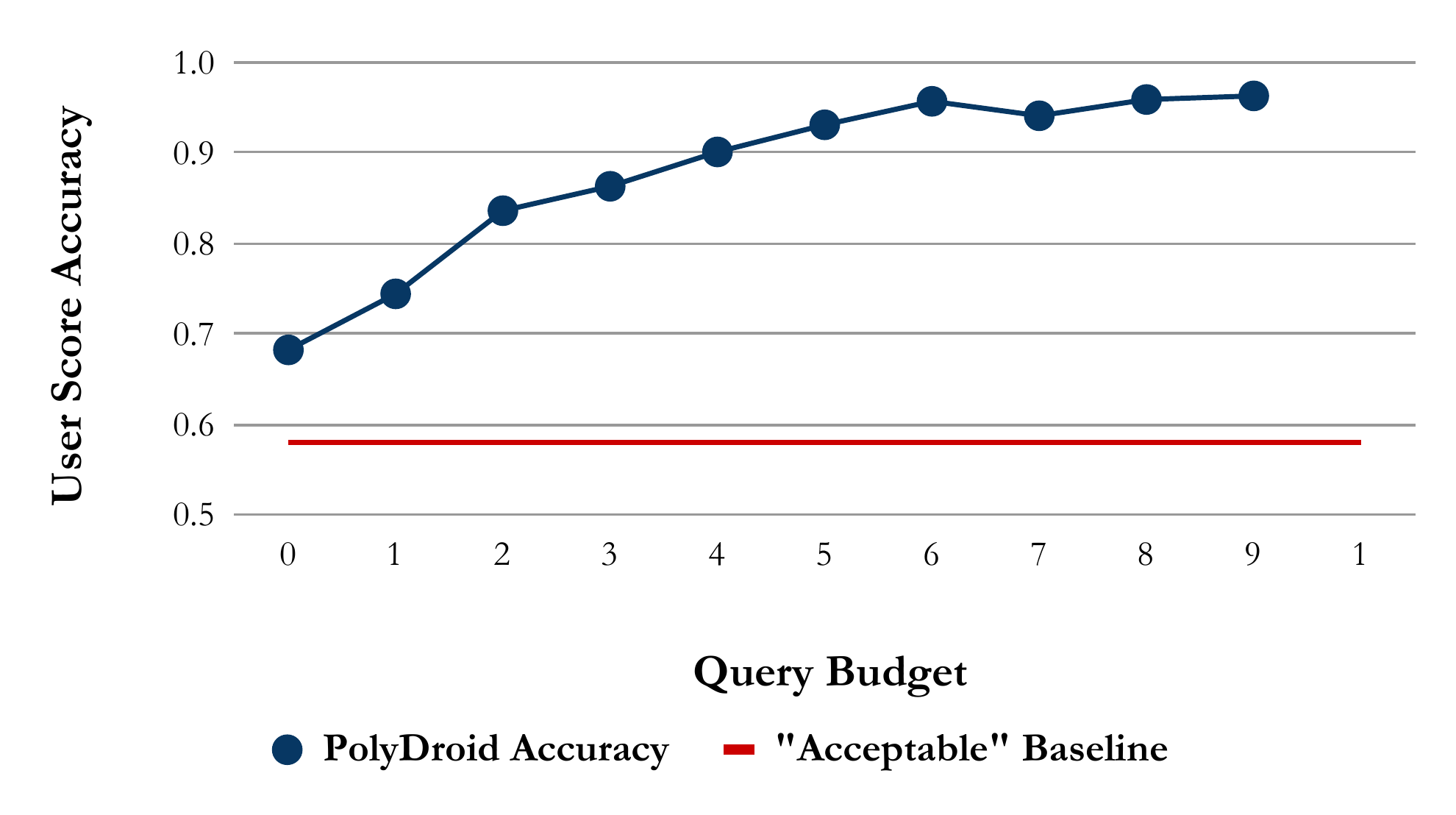}
\vspace{-0.2in}
\caption{Accuracy of \tool in predicting whether or not the average user score is at least 5 (neutral). Results are averaged across all 139 reductions of all 20 apps.}
\label{fig:prediction}
\vspace{-0.1in}
\end{figure}

\subsection{Resource Usage Estimation}
\label{sec:resourceconsumption}

We estimate the resource usage by running each reduction on a device and monitoring the resource consumption. Our implementation supports three kinds of resources: peak CPU usage, peak memory usage, and total network data usage. Our implementation runs each reduction in emulation, since modern Android emulators provide accurate device-specific resource usage estimates; if better accuracy is desired, the reductions can be run on a real device as well. We execute the test script that comes with each reduction, and then report the resource consumption for that run. More precisely, we ran the reduction using the Android Profiler within Android Studio. This tool outputs time series data for network traffic, and both the total memory allocated and the percent of CPU in use by the app we are analyzing. We use this data to compute the maximum CPU percentage, and maximum memory allocation, and total network data usage.

\subsection{Design Choices for Thompson Sampling}
\label{sec:thompsonimpl}

There are three design choices in our Thompson sampling algorithm: (i) the machine learning model $f(r;\theta)$, (ii) the choice of prior $p(\theta;\theta_0)$ on $\theta$, and (iii) the likelihood $\ell(\theta_0;\mathcal{H})$ used to choose the prior parameters $\theta_0$. We describe our choices below. Our choices correspond to using a Bayesian linear regression algorithm called automatic relevant determination (ARD) regression~\cite{bishop2006pattern}, both to estimate the prior parameters $\theta_0$ from historical data, and to compute the distribution $p(\theta;\theta_0,\mathcal{D})$.

\paragraph{{\bf Choice of model.}}

First, we choose the model to be
\begin{align*}
f(r;\theta)&=\langle\theta,\phi(r)\rangle+\epsilon_r \\
\epsilon_r&\sim\mathcal{N}(0,\sigma^2),
\end{align*}
for some $\sigma\in\mathbb{R}$. Here, $\phi:\mathcal{R}(a)\to\mathbb{R}^n$ is a feature mapping which maps each reduction $r\in\mathcal{R}(a)$ to a feature vector $\phi(r)\in\mathbb{R}^n$. The fifteen features used by \tool include ten metrics about the reduction (e.g. change in resolution), and five which describe the Activity where the reduction takes place (e.g. number and font size of TextViews, or blocks of text). 

\paragraph{{\bf Choice of prior.}}

Next, we choose the prior distribution over $\theta_a$ for the Thompson sampling algorithm to be
\begin{align*}
\theta_a\sim p(\cdot;\theta_0)=\mathcal{N}(\mu_0,\lambda\text{diag}(\sigma_0)^2),
\end{align*}
where the parameters of the (Gaussian) prior are $\theta_0=(\mu_0,\sigma_0)$, $\mu_0,\sigma_0\in\mathbb{R}^m$ encode the mean and variance the prior, $\text{diag}(\sigma_0)\in\mathbb{R}^{m\times m}$ is the diagonal matrix where the entries along the diagonal are given by the vector $\sigma_0$, and $\lambda\in\mathbb{R}$ is a hyperparameter. We chose $\lambda=20$ using cross-validation.

\paragraph{{\bf Choice of prior parameters.}}

Finally, we choose the prior parameters using the likelihood function
\begin{align*}
\ell(\theta_0;\mathcal{H})=p(\mathcal{H};\theta_0)=\prod_{(r,u^{(r)})\in\mathcal{H}}p(r,u^{(r)};\theta_0),
\end{align*}
where
\begin{align*}
p(r,u^{(r)};\theta_0)=\mathcal{N}(u^{(r)}-f(r;\theta_0),\sigma_1^2)
\end{align*}
and where $\sigma_1\in\mathbb{R}$ is a parameter; in ARD regression, $\sigma_1$ is estimated from data.

\begin{figure}
\includegraphics[width=\linewidth]{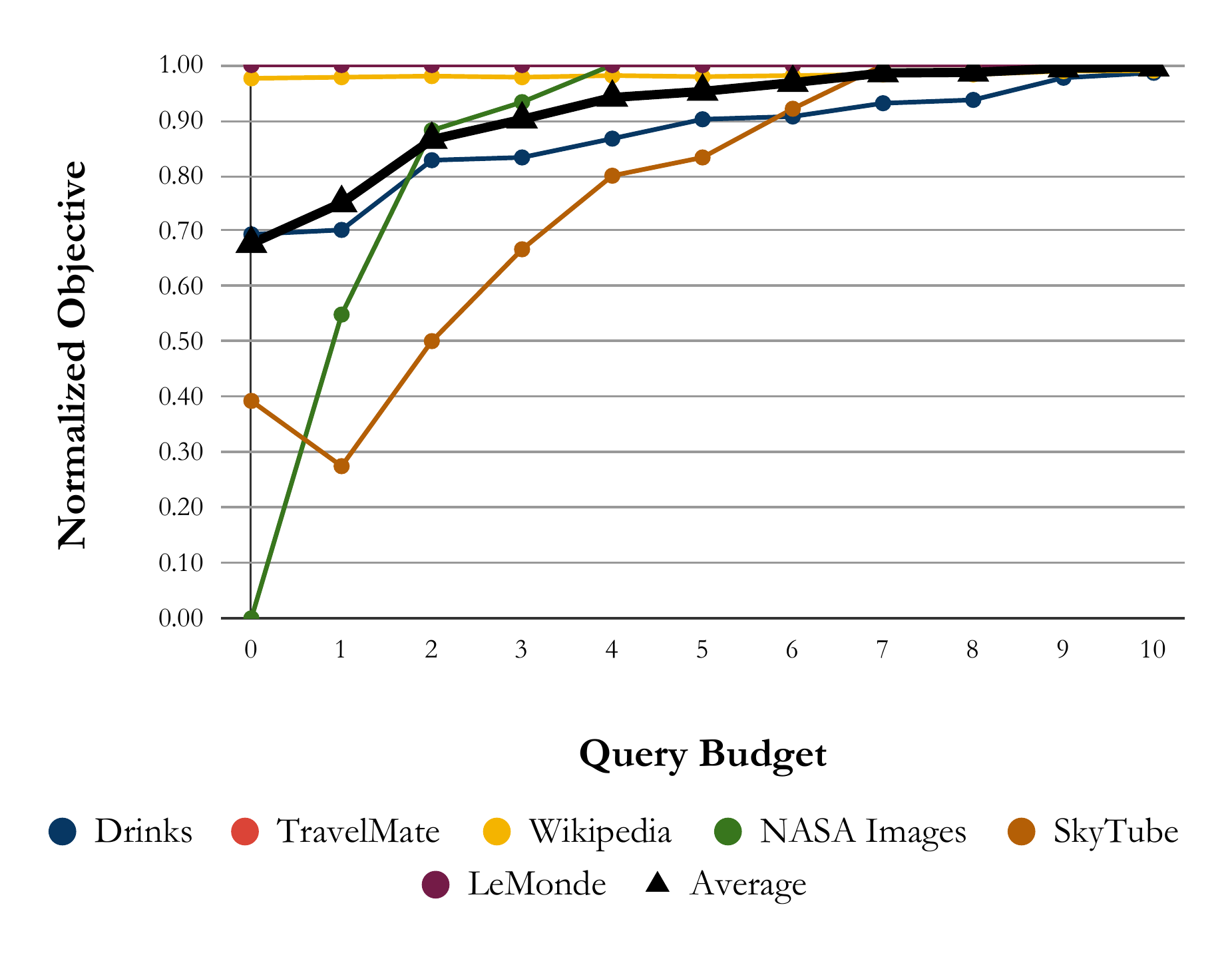}
\vspace{-0.3in}
\caption{Performance of \tool on finding good reductions for all six of our test apps. TravelMate and LeMonde achieved the optimal reduction using just the prior.}
\label{fig:optimization}
\vspace{-0.1in}
\end{figure}

\section{Empirical Evaluation}
\label{sec:eval}

We experimentally evaluate \tool by addressing the following research questions:
\begin{enumerate}
\item \textbf{Model effectiveness:} Can \tool accurately predict the user experience of different reductions?
\item \textbf{Active learning effectiveness:} Can \tool select good reductions based on a small fraction of user experiments?
\item \textbf{Personalization:} Can \tool be used to personalize reductions to individual end users?
\item \textbf{Robustness:} Is \tool effective across a range of different apps and different resource usage specifications?
\item \textbf{Need for personalized specifications:} Is it necessary to use different reductions for different specifications?
\end{enumerate}

\subsection{Experimental Setup}

\paragraph{{\bf Benchmark.}}

We collected 20 open-source Android apps. We used three criteria to choose which apps to analyze. First, we restricted to apps that could be modified to reduce consumption of at least one resource. Second, we restricted to actively maintained apps---i.e., apps with at least 50 commits and where the developers responded to issues for at least 90 days after initial publication. Finally, we selected apps from a range of categories (as categorized by the F-Droid~\cite{f-droid} open-source app store) to ensure that our findings are general. All apps were evaluated on a Google Nexus 5X emulator running Android 8.1 (Oreo). The emulator was allocated 4GB of hard disk storage and 1GB of RAM to ensure that there was sufficient space for each app to run without crashing.

Of the 20 apps, we chose six to be our test set. For each app $a$ in this test set, we use the remaining apps as historical data $\mathcal{H}$, and run Algorithm~\ref{alg:online} on $a$ to optimize the choice of reduction (for various choices of specifications). The six we selected were chosen to have a variety of possible modifications and drawn from multiple categories and with varying degrees of programmatic complexity. Over the six apps, there are 64 total potential versions that could contribute to the app's optimal configuration, or about 11 per app.

\paragraph{{\bf Reductions.}}

Google's Build for Billions~\cite{billions} initiative, which encourages more robust mobile development practices, identifies four key performance indicators for developers who desire to make their apps more accessible: handling mixed connectivity, building for device range, providing data controls, and using battery efficiently. We considered a number of modifications to the original apps, and concluded that reducing image resolution, altogether removing images, and eliminating transitions had the most significant impact across all of Google's key performance indicators. In total, we considered the seven possible modifications shown in Table~\ref{tbl:features}. These modifications can be applied independently to each Activity within a given Android app. Across the 20 apps in our benchmark, we constructed a total of 139 reductions based on these modifications.

\begin{figure}
\includegraphics[width=\linewidth]{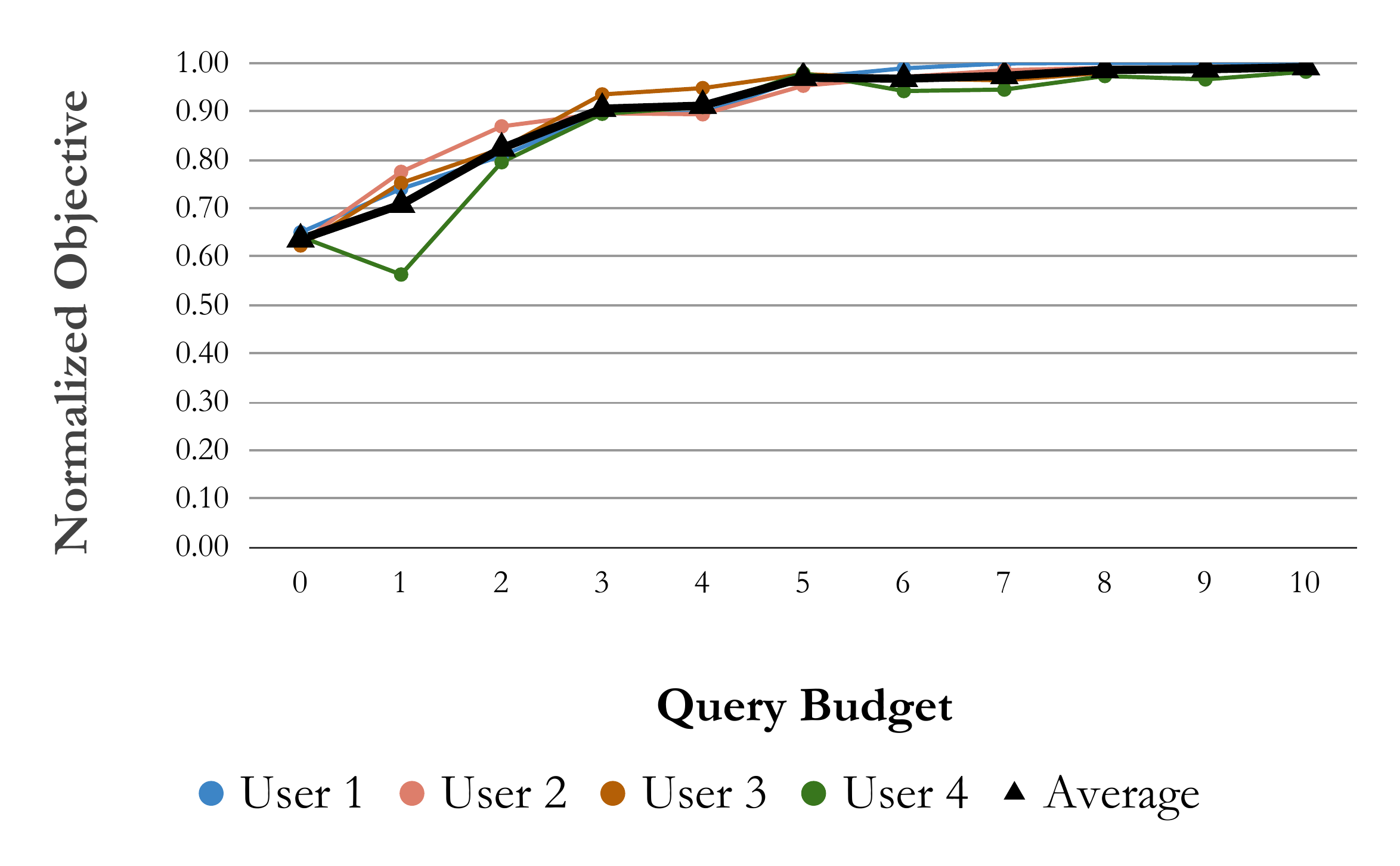}
\vspace{-0.25in}
\caption{Performance of \tool when optimizing for individual user preferences.}
\label{fig:personalize}
\vspace{-0.1in}
\end{figure}

\begin{figure*}
\includegraphics[width=\columnwidth, height=4.8cm]{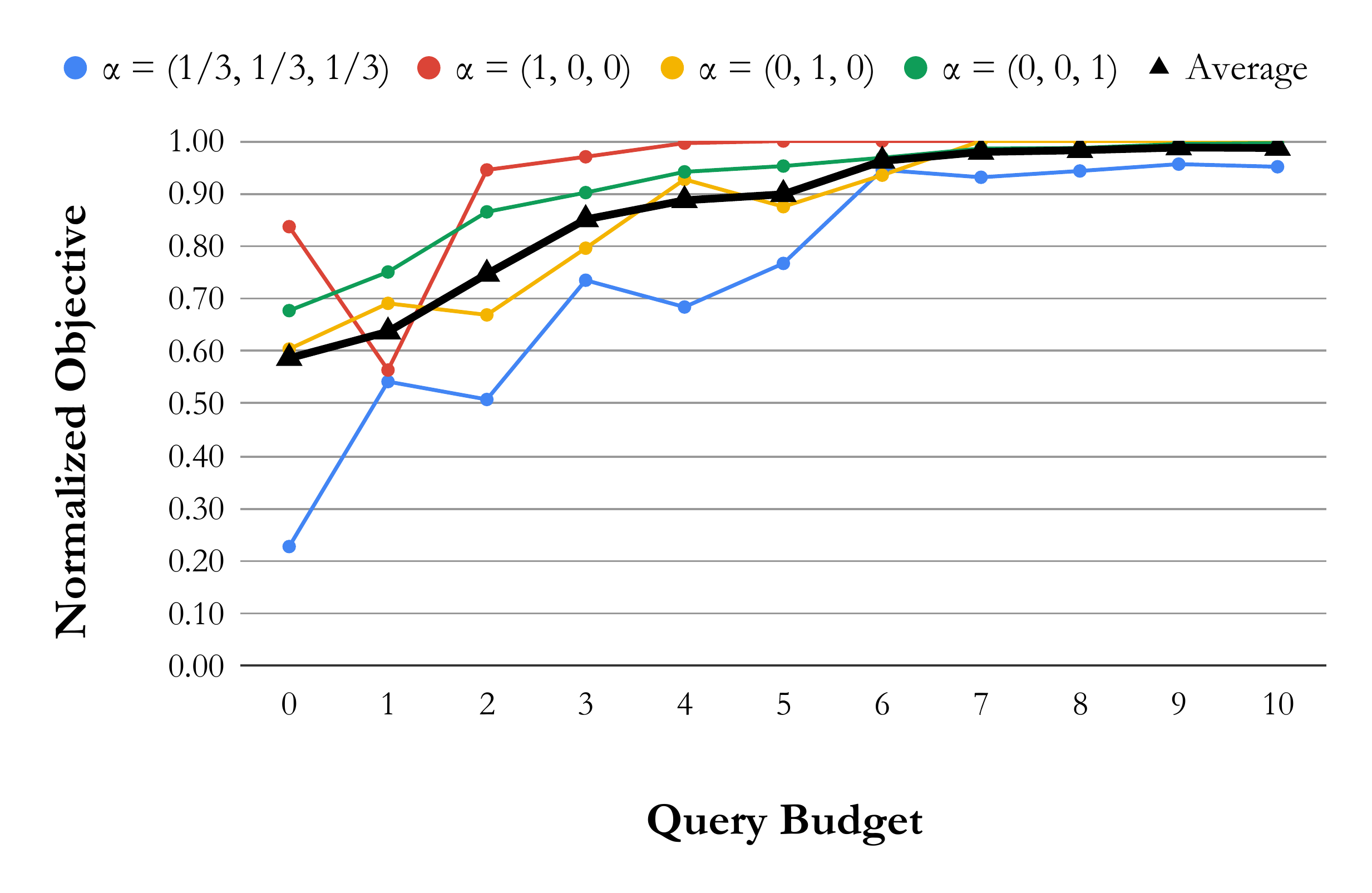}
\includegraphics[width=\columnwidth, height=4.8cm]{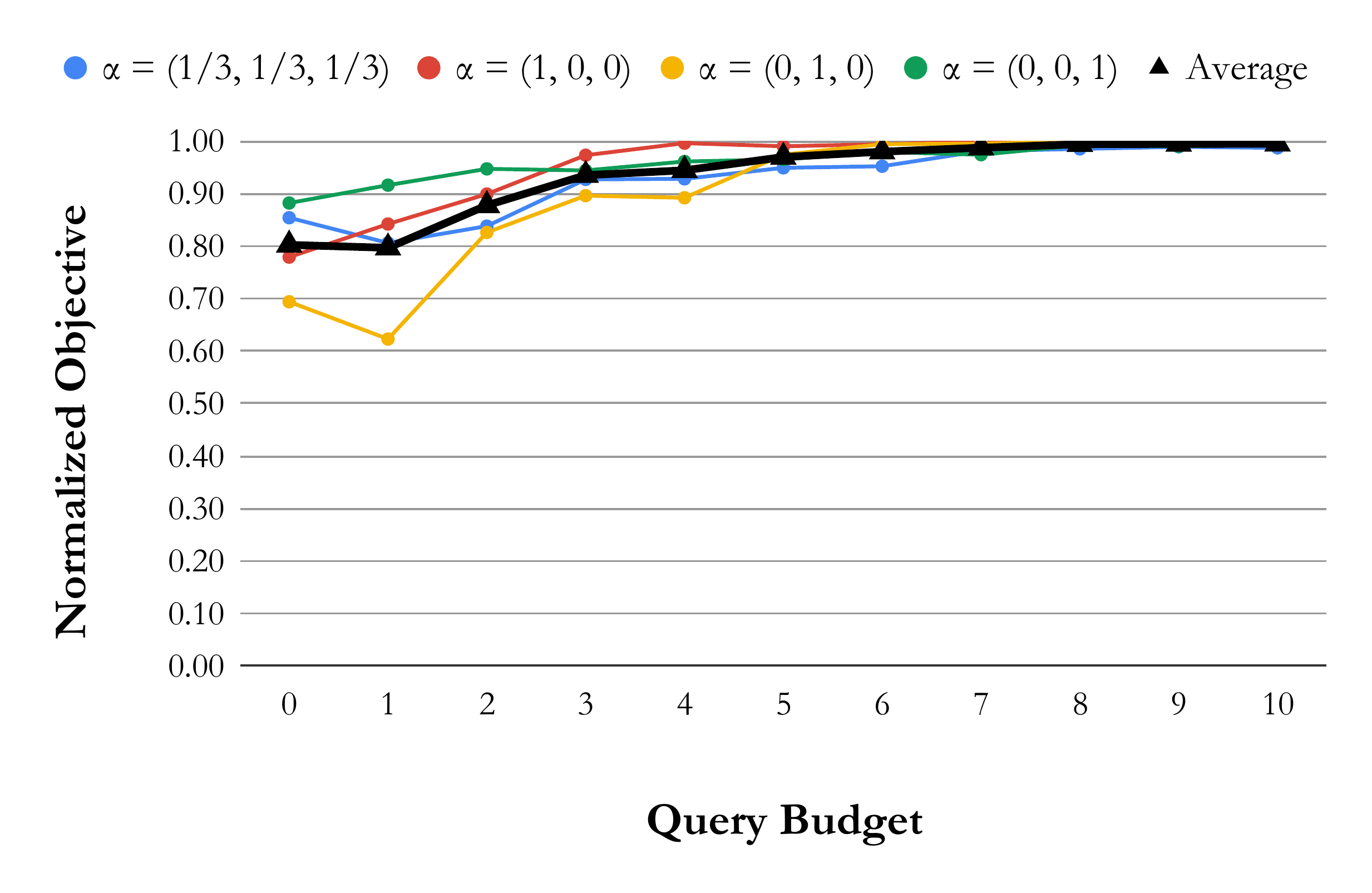}
\vspace{-0.15in}
\caption{Performance of \tool for different specifications, with $\lambda = 1$ (left) and $\lambda = 3$ (right).}
\label{fig:robust}
\end{figure*}

\paragraph{{\bf Test scripts.}}

Recall that we determine the user experience score $u^{(r)}$ using the approach described in Section~\ref{sec:userqueries} in conjunction with test scripts for each reduction and for the original app. We manually implemented test scripts to capture the Activity in which the modification took place. For image modifications (i.e., reduced resolution or removal), participants saw complete examples of the images that were changed. For transition removal modifications, participants saw a button click that executes the animation and then switches activities.

\paragraph{{\bf Objective function.}}

Since we consider three resources (CPU usage, memory usage, network data usage), the user-provided resource usage specification has the form $(\lambda,\alpha)$, where $\lambda\in\mathbb{R}$ and $\alpha\in\mathbb{R}^3$. For our evaluation, we use the notation $\alpha_{\text{CPU}}$, $\alpha_{\text{mem}}$, and $\alpha_{\text{net}}$. Similarly, given a reduction $r$, we use $w_{\text{CPU}}^{(r)}$ to denote its CPU usage savings, $w_{\text{mem}}^{(r)}$ to denote its memory usage savings, and $w_{\text{net}}^{(r)}$ to denote its network data usage savings (all normalized with respect to the original app). Thus, our objective is
\begin{align*}
J(r;\lambda,\alpha) = u^{(r)} + \lambda\left(\alpha_{\text{CPU}}w_{\text{CPU}}^{(r)} + \alpha_{\text{mem}}w_{\text{mem}}^{(r)} + \alpha_{\text{net}}w_{\text{net}}^{(r)}\right).
\end{align*}

\subsection{Can \tool Accurately Predict User Experience?}

We begin by demonstrating that \tool can accurately predict the user experience score of a reduction $r$. In particular, we study how active learning can be used to effectively predict user experience using just a few user experiments. For this experiment, we binarize the labels based on whether $u^{(r)}\ge5$ (i.e., a neutral score according to our survey).

We find that with zero actively chosen samples, \tool already has accuracy 69.0\%. In contrast, random predictions have accuracy 50.0\%, and always predicting ``acceptable'' (i.e., score $\ge5$) has accuracy 59.0\%. Next, we evaluate the accuracy of \tool given a small query budget $B$. For this task, we use the specification $(\lambda = 1, \alpha_{\text{CPU}} = 1/3, \alpha_{\text{mem}} = 1/3, \alpha_{\text{net}} = 1/3)$ to actively select reductions to label. Figure~\ref{fig:prediction} depicts our findings. As can be seen, after two queries, the accuracy of \tool improves to over 80\%, and after four queries, the accuracy improves to over 90\%.

\subsection{Can \tool Find Near-Optimal Reductions?}
\label{sec:evaloptimal}

Next, we test whether \tool can effectively leverage its ability to predict user experience scores to help find near-optimal reductions of our six test apps. For each reduction, we calculate the average of the objective $J(\hat{r};\lambda,\alpha)$, where $\hat{r}$ is the reduction returned by Algorithm~\ref{alg:online}. We average our results over 25 runs (due to the randomness in the Thompson sampling algorithm). For this experiment, we use $(\lambda=1, \alpha_{\text{CPU}}=0, \alpha_{\text{mem}}=0, \alpha_{\text{net}}=1)$---i.e., an equal tradeoff between user experience and network data savings. We believe this specification is a commonly desired choice; we explore other specifications in Section~\ref{sec:evalrobust}. We normalize the objective value achieved by the reduction $\hat{r}$ returned by \tool as follows:
\begin{align*}
\rho(r;\lambda,\alpha)=\frac{J(r;\lambda,\alpha)-J(a;\lambda,\alpha)}{J(r^*;\lambda,\alpha)-J(a;\lambda,\alpha)}.
\end{align*}
In particular, $\rho(a;\lambda,\alpha)=0$ (i.e., choosing the original app achieves a score of 0), and $\rho(r^*;\lambda,\alpha)=1$ (i.e., choosing the optimal reduction achieves a score of 1); note that $\rho(r;\lambda,\alpha)$ can be negative.

In Figure~\ref{fig:optimization}, we show the normalized objective value $\rho(\hat{r};\lambda,\alpha)$ achieved by \tool as a function of the query budget $B$. As can be seen, without any queries (i.e., based only on the historical data), \tool achieves 67.7\% of the optimal score. Querying just two reductions improves performance, on average, to over 80\% of the optimal score. Thus, \tool requires very few queries to obtain a reduction that is close in quality to the optimal reduction for the given resource usage specification.

\subsection{Can \tool Personalize Reductions?}

Next, we demonstrate that \tool can personalize reductions based on the user experience preferences of individual users. We take the scores of four individuals who completed all surveys, and run \tool using the scores they assigned as labels, focusing on the specification $(\lambda=0.5, \alpha_{\text{CPU}}=0, \alpha_{\text{mem}}=0, \alpha_{\text{net}}=1)$. We average our results over 25 runs. For each run, we select a test app at random, and run Algorithm~\ref{alg:online} to optimize the reduction.

In Figure~\ref{fig:personalize}, we show the normalized objective value $\rho(\hat{r};\lambda,\alpha)$ achieved as a function of the query budget $B$. As can be seen, \tool achieves, on average, 80\% of the optimal score using just 2 queries, and 90\% of the optimal score using just 3 queries. Importantly, despite the fact that individual survey responses have higher variance, we achieve results that are similar to those in Section~\ref{sec:evaloptimal} (which are for average user scores).

\begin{figure*}
\begin{tabular}{ccc}
\includegraphics[width=0.3\linewidth]{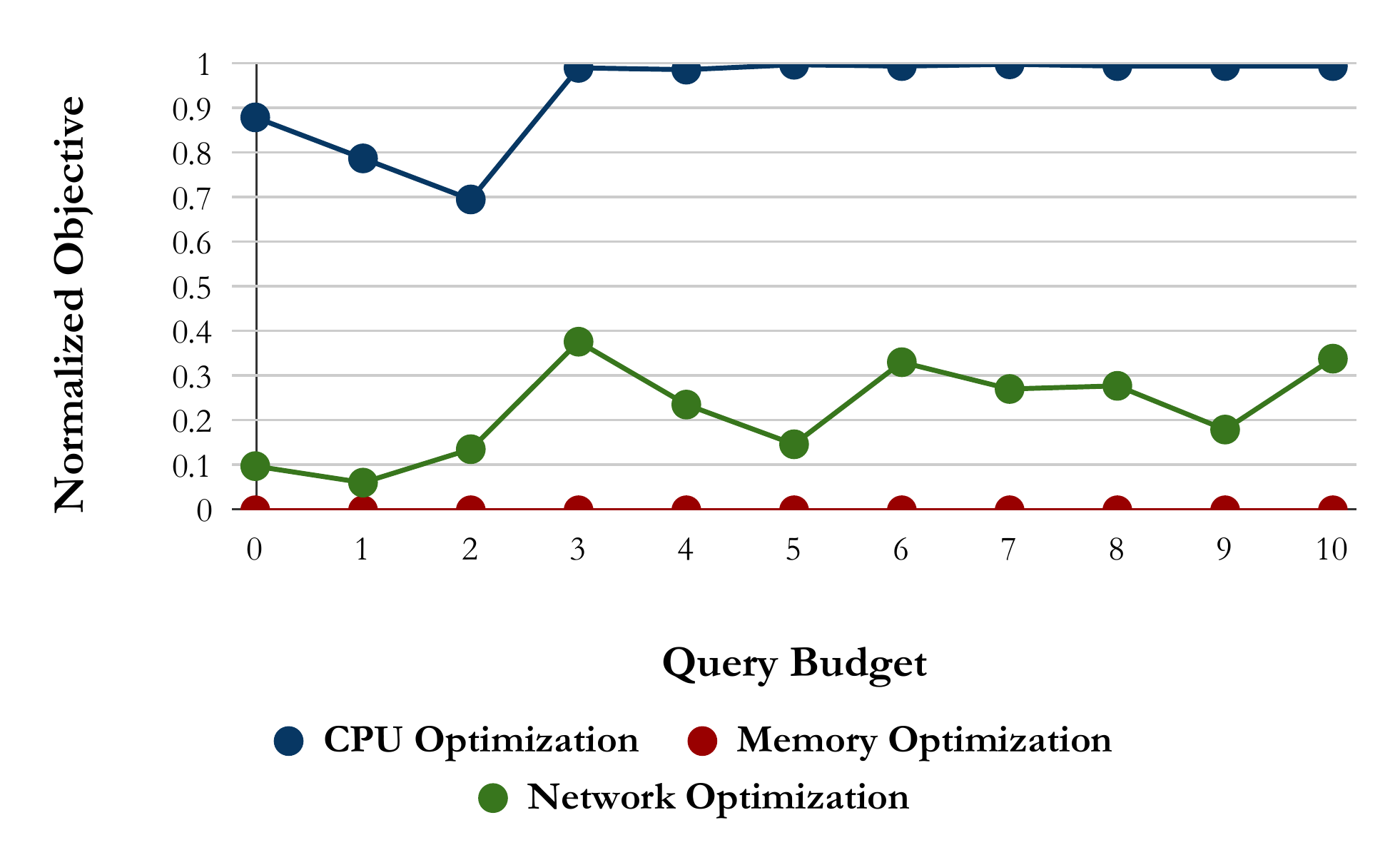} &
\includegraphics[width=0.3\linewidth]{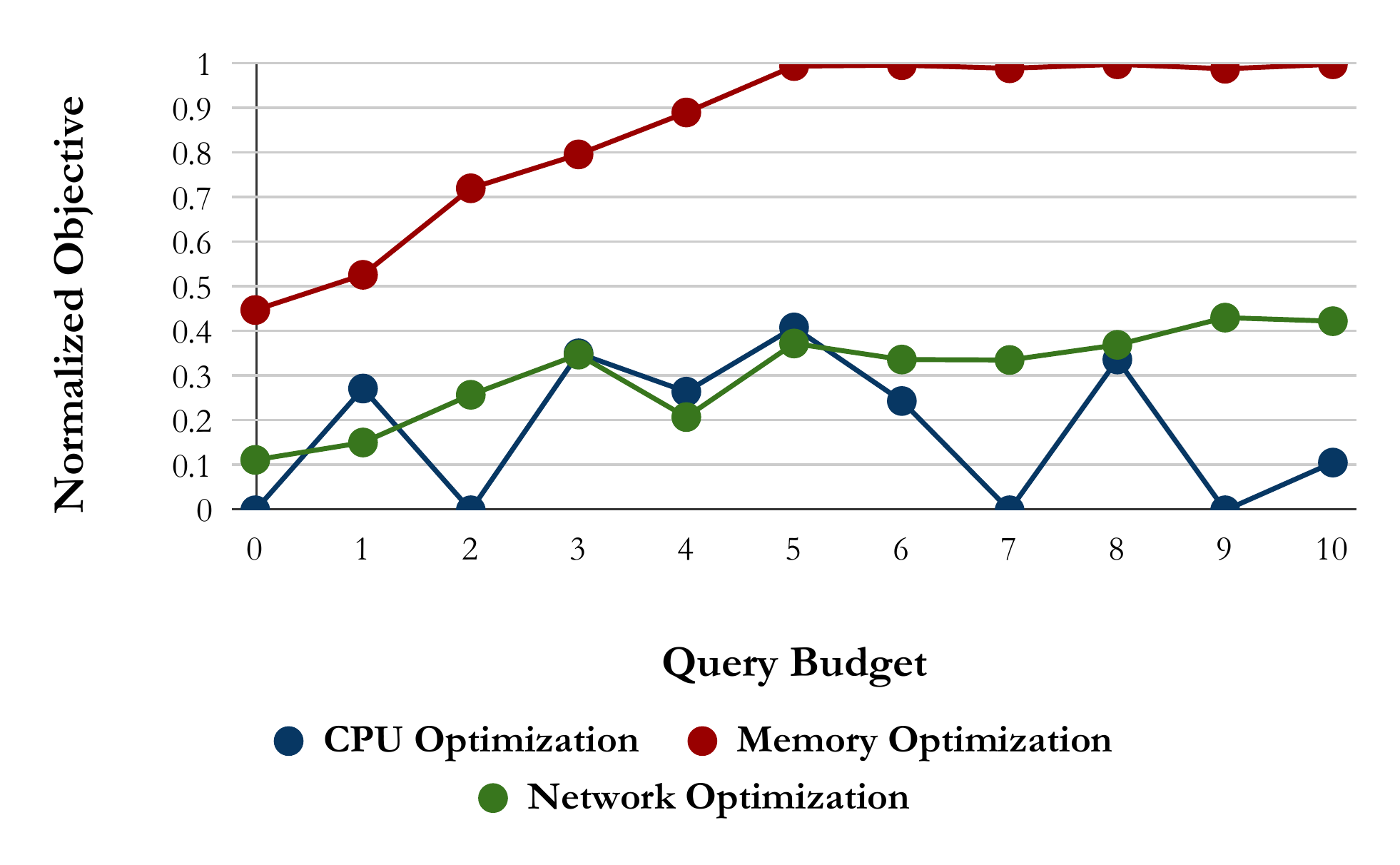} &
\includegraphics[width=0.3\linewidth]{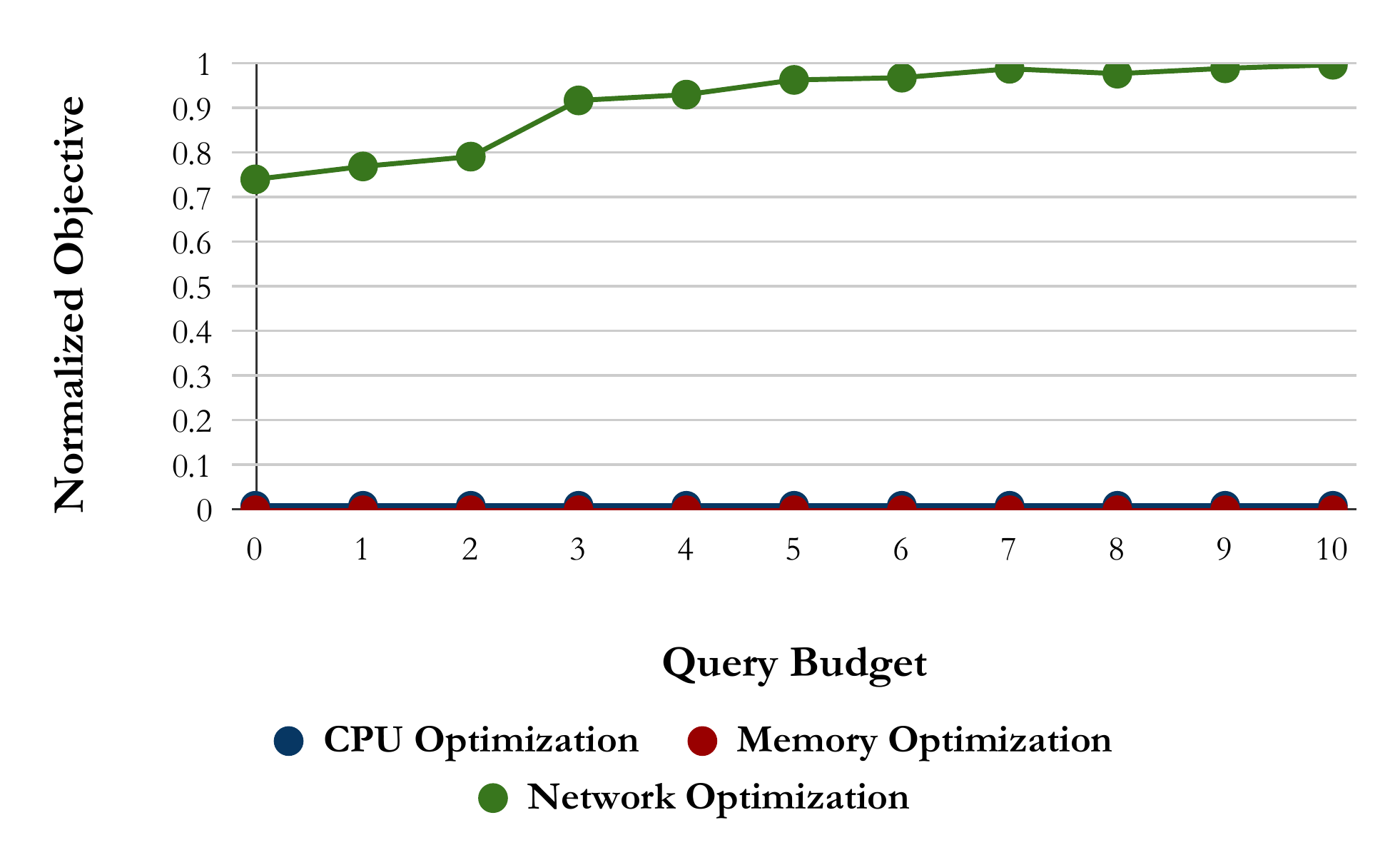} \\
(a) & (b) & (c)
\end{tabular}
\vspace{-0.15in}
\caption{Performance of \tool when optimizing (a) CPU usage, (b) memory usage, and (c) network data usage.}
\vspace{-0.1in}
\label{fig:stratify}
\end{figure*}

\subsection{Does \tool Work for Different Specifications?}
\label{sec:evalrobust}

Next, we study how well \tool works for a range of different resource usage specifications. We test across eight specifications, including two choices $\lambda\in\{1,3\}$ and four choices $\alpha\in\{(1,0,0),(0,1,0),(0,0,1),(1/3,1/3,1/3)\}$. In other words, the choices of $\alpha$ either weight each of resources equally, or fully weight one specific resource in CPU, memory, or network data. The first choice models the scenario where the end user simply needs a downgraded app, and the remaining three choices model scenarios where the end user wants to conserve a specific resource.

In Figure~\ref{fig:robust}, we show the normalized objective value $\rho(\hat{r};\lambda,\alpha)$ achieved as a function of the query budget. \tool successfully leverages its queries in all cases to improve performance. In particular, in all cases, \tool achieves 85\% of the optimal score after querying 4 of all possible reductions.

\subsection{Is Personalization Necessary?}

Next, we study whether \tool actually needs to find different reductions for different resource usage specifications. Conceivably, one optimal reduction of an app could satisfy all possible resource constraints, making specifications unnecessary. We consider three specifications with $\lambda=3$ and $\alpha\in\{(1,0,0),(0,1,0),(0,0,1)\}$. In other words, these specifications optimize a single resource.

In Figure~\ref{fig:stratify}, we show the results for (a) the specification $(1,0,0)$ (i.e., optimizing CPU usage), (b) $(0,1,0)$ (i.e., optimizing memory usage), and (c) $(0,0,1)$ (i.e., optimizing network data usage). In each plot, we show the normalized objective value $\rho(\hat{r};\lambda,\alpha)$ achieved for each of the three specifications. These plots show that if \tool is used to optimize one resource, then the other resources are not necessarily optimized, and often even perform very poorly. Therefore, \tool must select different reductions for different specifications.

\subsection{Threats to Validity}
\label{subsec:ttv}

There are several threats to the validity of our studies.
The main threat to internal validity arises because \tool uses an emulator to
estimate resource consumption as well as to survey user experience.
The resource estimates and user ratings should ideally be obtained from
the user's device, since they could differ from the emulator-based results.
While current Android emulation technology is quite sophisticated,
such discrepancies could increase the number of user experiments needed by
\tool's active learning framework to predict the optimal reduction.
Another threat arises because the reductions that we evaluated (Table \ref{tbl:features})
were not designed by the original app developers. While our choice of reductions
was informed by the guidelines in Google's Build for Billions initiative,
individual app developers could choose reductions different from ours.
Such custom reductions could prevent \tool from effectively leveraging prior knowledge
about other apps.

Threats to external validity arise when the results of the
experiment cannot be generalized. We evaluated \tool using only 20 apps. 
Thus, the performance of our technique may vary for other apps. 
However, our apps are representative of typical Android apps considering the
problem that our technique addresses.

\vspace{-0.05in}
\section{Related Work}
\label{sec:related}

Our work is related to existing work on approximate computing \cite{misailovic:qoe,hoffmann:qoe,park:qoe,sidiroglou-douskos:qoe,sampson:qoe,misailovic:chisel}.  The most closely related to ours is Park et al. \cite{axgames}'s solution to understanding the correlation between displayed
quality loss and user acceptability.  They create games designed to evaluate a user's response to quality tradeoff, and employ crowdsourcing through Amazon's Mechanical Turk to come to a reliable conclusion as to the degree of quality loss which is still acceptable.  \tool also uses uses crowdsourcing to understand optimal reductions, but differs in that it uses active learning to reduce user experiments and is able to personalize reductions to individual end users. 

Canino et al. \cite{canino2018fse} propose an energy optimization framework, Aeneas, which provides a solution to application-level energy management through the use of stochastic optimization.  Their API allows developers to specify knobs for energy optimizations within source code, and leaves it up to the process of stochastic optimization to find the appropriate configuration for these knobs to save energy.  \tool is similar in making resource savings a primary goal, but solves the problem by directly querying users on what resource-saving configuration would still be acceptable.

There has also been work on understanding the quality of experience with respect to mobile apps \cite{mao:qoe}; Chen et al. \cite{qoedoctor} propose a tool which detects app quality problems within both the system and network stack through analysis of interaction test scripts.  Whereas their tool attempts to programatically identify a more suitable configuration, \tool uses both crowdsourcing and the preferences of the end user to find the suitable configuration.  

To counter the trend of increasing software complexity and bloat, 
a recent body of work has proposed program debloating \cite{Rastogi:2017,Jiang:2016,reddroid,proguard,Quach:2018,Bhattacharya:2013,Heo:CCS18},
which concerns techniques to identify and remove undesired functionality. 
RedDroid \cite{reddroid} and ProGuard \cite{proguard} target debloating Android apps in order to reduce size and improve performance.
For instance, ProGuard can make Android apps up to 90\% smaller and up to 20\% faster.
Chisel \cite{Heo:CCS18} targets debloating general-purpose software but requires the programmer to provide a test script to guide the reduction process
and does not provide any safety guarantees.
Bhattacharya et al. \cite{Bhattacharya:2013} propose an analysis technique to detect possible sources of bloat
in Java programs when optional features are no longer required.
A user must intervene for confirming the detected statements and removing them.

There is a large body of research on detecting and reducing runtime memory
bloat~\cite{androidgo, Nguyen:2018, Xu:2010software, Xu:2010detecting, Xu:2014, Xu:2009}.
Android Go \cite{androidgo} is a slimmed down version of the Android OS
which aims to run the OS successfully on entry-level phones with RAM ranging from
512 MB to 1GB. Apps on this platform occupy about 50\% less space and run almost 15\%
faster than their counterparts on the regular platform.
These works are complementary to \tool's approach as they aim to improve resource
utilization without sacrificing user experience.

\vspace{-0.05in}
\section{Conclusion}
\label{sec:conclude}

We presented \tool, an active learning driven framework for optimizing mobile applications based on user preferences in resource constrained environments.  \tool offers an effective solution to accomodating the ever-expanding variety of smartphone configurations and the various environments in which they are used, all the while tailoring to the preferences of the end user. Active learning allows our framework to understand a new app and, as expirimental results show, determine an accurate reduction configuration for each individual end user's preferences, even though it used manyfold less queries than the traditional approach.  

In the future, we plan on extending \tool in many different ways.  Our current work focuses on removing or reducing images and transitions, but in the future we plan to explore other reductions such as batching together network calls and releasing them in timed intervals, and disabling expensive background tasks.  Furthermore, we plan to build a web service for developers to use our tool, as well as expand to the iOS environment.

\balance
\bibliographystyle{ACM-Reference-Format}
\bibliography{citations}

\end{document}